\documentclass[12pt]{article}
\usepackage{amsmath}
\usepackage{amssymb}
\usepackage{amsfonts}
\usepackage{graphics}

\newcommand{\field}[1]{\mathbb{#1}}
\newcommand{\R}{\field{R}}

\title{Addendum to ``An update on the classical and quantum harmonic oscillators on the sphere and the hyperbolic plane in polar coordinates''}

\author{C.\ Quesne\thanks{E-mail address: cquesne@ulb.ac.be} \\
{\small\sl Physique Nucl\'eaire Th\'eorique et Physique Math\'ematique, 
Universit\'e Libre de Bruxelles,} \\ 
{\small \sl Campus de la Plaine CP229, Boulevard~du Triomphe, B-1050
Brussels, Belgium}}
\date{ }
\begin{document}
\baselineskip=22pt plus 1pt minus 1pt
\maketitle

\begin{abstract}
The classical and quantum solutions of a nonlinear model describing harmonic oscillators on the sphere and the hyperbolic plane, derived in polar coordinates in a recent paper [Phys.\ Lett.\ A 379 (2015) 1589], are extended by the inclusion of an isotonic term.
\end{abstract}

\vspace{0.5cm}


 
\newpage
%
%
In a recent paper \cite{cq}, we presented a simple derivation in polar coordinates of the classical solutions of a nonlinear model describing harmonic oscillators on the sphere and the hyperbolic plane, previously derived in cartesian coordinates \cite{carinena04}, and we identified the nature of the classical orthogonal polynomials entering the bound-state radial wavefunctions of the corresponding quantum model \cite{carinena07}. One of the interests of such a model is that it provides a two-dimensional generalization of the classical nonlinear oscillator introduced by Mathews and Lakshmanan as a one-dimensional analogue of some quantum field theoretical models \cite{mathews, lakshmanan}.\par
%
%
The purpose of the present addendum is to show that with some small changes both the classical and quantum results of \cite{cq} can be adapted to deal with a two-dimensional extension of a recent study of a nonlinear oscillator with an isotonic term performed in one dimension \cite{ranada}.\par
%
%
In polar coordinates $r$, $\varphi$, the Lagrangian of the classical harmonic oscillator with an isotonic term reads
\begin{equation}
  L = \frac{1}{2} \left(\frac{\dot{r}^2}{1 + \lambda r^2} + \frac{J^2}{r^2}\right) - \frac{1}{2} \frac{\alpha^2 
  r^2}{1 + \lambda r^2} - \frac{k}{2r^2},  \label{eq:L} 
\end{equation}
where $\alpha$ and $k$ are some real, positive constants, $J = r^2 \dot{\varphi}$ denotes the angular momentum, which is a constant of the motion, and the nonlinearity parameter $\lambda$ is related to the curvature $\kappa$ by $\lambda = - \kappa$, with $\kappa > 0$ for a sphere and $\kappa < 0$ for a hyperbolic plane.\par
%
%
The solutions of the Euler-Lagrange equations corresponding to (\ref{eq:L}) are obtained by successively integrating the differential equation
\begin{equation}
  2dt = \frac{dr^2}{\sqrt{a + br^2 + cr^4}}, \qquad a = -J^2 -k, \qquad b = C + \frac{\alpha^2}{\lambda} - 
  \lambda (J^2+k), \qquad c = C\lambda,  \label{eq:diff}
\end{equation}
inverting the resulting solutions $t = t(r^2)$ to yield $r^2 = r^2(t)$, and finally integrating the differential equation $\dot{\varphi} = J/r^2(t)$. In (\ref{eq:diff}), $C$ denotes some integration constant, which, as before, can be related to the energy through 
\begin{equation}
  E = \frac{1}{2} C + \frac{\alpha^2}{2\lambda} \qquad \text{or} \qquad C = 2E - \frac{\alpha^2}{\lambda}.
\end{equation}
\par
%
%
On the other hand, the energy can be written as
\begin{equation}
  E = \frac{1}{2} \frac{\dot{r}^2}{1 + \lambda r^2} + V_{\rm eff}(r), \qquad V_{\rm eff}(r) = \frac{1}{2}
  \frac{\alpha^2 r^2}{1 + \lambda r^2} + \frac{J^2+k}{2r^2},
\end{equation}
where the isotonic term $k/(2r^2)$ gives an additional contribution to the effective potential $V_{\rm eff}(r)$. Due to this, the latter always goes to $+\infty$ for $r \to 0$, so that there is no need to distinguish between $J=0$ and $J\ne 0$ as in \cite{cq} (except for the integration of the angular differential equation). Moreover, the effective potential, which still goes to $\alpha^2/(2\lambda)$ for $r \to \infty$ if $\lambda > 0$ or to $+\infty$ for $r \to 1/\sqrt{|\lambda|}$ if $\lambda < 0$, has now a minimum $V_{\rm eff, min} = \frac{1}{2} \sqrt{J^2+k} \left(2\alpha - \lambda \sqrt{J^2+k}\right)$ at $r_{\rm min} = \left[\sqrt{J^2+k}/\left(\alpha - \lambda \sqrt{J^2+k}\right)\right]^{1/2} \in (0, +\infty)$ or $\left(0, 1/\sqrt{|\lambda|}\right)$ (according to which case applies). For $\lambda > 0$, such a minimum only exists for $J$ values such that $\sqrt{J^2+k} < \alpha/\lambda$, which implies that $k$ must be such that $k < \alpha^2/\lambda^2$. This shows that bounded trajectories are then limited to low angular momentum values and weak isotonic term.\par
%
%
{}For $\lambda>0$ and $V_{\rm eff,min} < E < \alpha^2/(2\lambda)$ or $\lambda<0$, the complete solution is given by
\begin{equation}
\begin{split}
  & r^2 = A \sin(2\omega t + \phi) + B, \quad B-A \le r^2 \le B+A, \\
  & A = \frac{1}{2|\lambda|\omega^2} \sqrt{\left[\left(\alpha - \lambda \sqrt{J^2+k}\right)^2 - \omega^2
      \right]\left[\left(\alpha + \lambda \sqrt{J^2+k}\right)^2 - \omega^2\right]}, \\
  & B = \frac{\alpha^2 - \lambda^2 (J^2+k) - \omega^2}{2\lambda\omega^2}, \quad \phi \in [0, 2\pi), \\
  & \omega = \sqrt{|c|}, \quad E = \frac{\alpha^2 - \omega^2}{2\lambda}, \\
  & \tan\left(\frac{\sqrt{J^2+k}}{J}(\varphi - K)\right) = \frac{\omega}{\sqrt{J^2+k}} \left[B 
       \tan\left(\omega t + \frac{\phi}{2}\right) + A\right] \quad \text{if $J\ne 0$}, \\
  & \varphi = K \quad \text{if $J=0$},
\end{split}
\end{equation}
and describes bounded trajectories.\par
%
%
{}For $\lambda>0$ and $\alpha^2/(2\lambda) < E < +\infty$, the trajectories are unbounded and characterized by
\begin{equation}
\begin{split}
  & r^2 = A \cosh(2\omega t + \phi) + B, \quad A+B \le r^2 < +\infty, \\
  & A = \frac{1}{2\lambda\omega^2} \sqrt{\left[\left(\alpha - \lambda \sqrt{J^2+k}\right)^2 + \omega^2
      \right]\left[\left(\alpha + \lambda \sqrt{J^2+k}\right)^2 + \omega^2\right]},  \\
  & B = - \frac{\alpha^2 - \lambda^2 (J^2+k) + \omega^2}{2\lambda\omega^2}, \quad
      \phi \in \R, \\
  & \omega = \sqrt{c}, \quad E = \frac{\alpha^2 + \omega^2}{2\lambda}, \\
  & \tan\left(\frac{\sqrt{J^2+k}}{J} (\varphi - K)\right) = \frac{\omega}{\sqrt{J^2+k}} (A-B) 
      \tanh\left(\omega t + \frac{\phi}{2}\right) \quad \text{if $J\ne 0$}, \\
  & \varphi = K \quad \text{if $J=0$}.
\end{split}  
\end{equation}
\par
%
%
{}Finally, for $\lambda>0$ and $E = \alpha^2/(2\lambda)$, we get a limiting unbounded trajectory, specified by
\begin{equation}
\begin{split}
  & r^2 = (At + \phi)^2 + B, \quad B \le r^2 < +\infty, \\
  & A = \sqrt{\frac{1}{\lambda}[\alpha^2 - \lambda^2 (J^2+k)]}, \quad B = \frac{\lambda (J^2+k)}{\alpha^2 -
      \lambda^2 (J^2+k)}, \quad \phi \in \R, \\
  & \tan\left(\frac{\sqrt{J^2+k}}{J}(\varphi - K)\right) = \frac{A}{\sqrt{J^2+k}} (At + \phi) \quad 
      \text{if $J\ne 0$}, \\
  & \varphi = K \quad \text{if $J=0$}.
\end{split}  \label{eq:sol-3}
\end{equation}
\par
%
%
Turning now ourselves to the corresponding quantum problem, we note that the Schr\"odinger equation of \cite{cq} becomes
\begin{equation}
  \left((1+\lambda r^2) \frac{\partial}{\partial r^2} + (1+2\lambda r^2) \frac{1}{r} \frac{\partial}{\partial r}
  + \frac{1}{r^2} \frac{\partial^2}{\partial\varphi^2} - \frac{\beta(\beta+\lambda) r^2}{1+\lambda r^2}
  - \frac{k}{r^2}+ 2E\right) \Psi(r,\varphi) = 0,
\end{equation}
where $\hbar=1$ and $\alpha^2 = \beta(\beta + \lambda)$ as before. After separating the variables $r$ and $\varphi$ by setting $\Psi(r,\varphi) = R(r) e^{{\rm i}m\varphi}/\sqrt{2\pi}$, where $m$ may be any positive or negative integer or zero, we get the radial equation
\begin{equation}
  r^2 (1+\lambda r^2) R'' + r (1+2\lambda r^2) R' + \left(- \frac{\beta(\beta+\lambda) r^4}{1+\lambda r^2}
  + 2Er^2 - \mu^2\right) R = 0,  \label{eq:radial}
\end{equation}
where $\mu^2 = m^2 + k$ and $\mu$ is defined as the positive square root $\sqrt{m^2+k}$.\par
%
%
The solutions of Eq.~(\ref{eq:radial}) are given by
\begin{equation}
\begin{split}
  & R_{n_r,\mu}(r) \propto (1+\lambda r^2)^{-\beta/(2\lambda)} r^{\mu} P^{\left(\mu, -\frac{\beta}{\lambda}
      - \frac{1}{2}\right)}_{n_r}(1+2\lambda r^2), \\
  & E_n = (n+1) \left(- \frac{\lambda}{2} n + \beta\right), \quad n = 2n_r + \mu,
\end{split}
\end{equation}
where $n_r= 0$, 1, 2,\ldots, but the values taken by $n$ are not necessarily integer anymore. Normalizable radial wavefunctions with respect to the measure $(1+\lambda r^2)^{-1/2} r dr$ on the interval $(0, +\infty)$ if $\lambda > 0$ or $(0, 1/\sqrt{|\lambda|})$ if $\lambda < 0$ correspond to all possible values of $n_r$ and $m$ in the latter case, but are restricted by the condition $n < \frac{\beta}{\lambda} - \frac{1}{2}$ in the former.\par
%
%

\end{document}